\documentclass[english,aps,prl,twocolumn,letterpaper]{revtex4}

\usepackage[T1]{fontenc}

\usepackage{graphicx}
\usepackage{epstopdf}
\usepackage{amssymb}
\usepackage{amsmath}
\usepackage{tikz}
\usetikzlibrary{decorations.pathmorphing}
\usetikzlibrary{arrows,shapes}
\usepackage{subfigure}
\usepackage{breakurl}
\usepackage[breaklinks=true,urlcolor=blue,hyperindex,colorlinks,bookmarks=true,linkcolor=black, citecolor=black]{hyperref}

\usepackage[normalem]{ulem}
\usepackage[abs]{overpic}
\usepackage{color}
\usepackage{rotating}
\usepackage{enumerate}

\newcommand{\ket}[1]{\left|#1\right\rangle}
\newcommand{\bra}[1]{\left\langle#1\right|}
\newcommand{\braket}[2]{\left\langle#1\right|\left.#2\right\rangle}

\begin{document}

\title{On the physical realizability of quantum stochastic walks}

\author{Bruno G. Taketani}
\affiliation{Theoretical Physics, Saarland University, Campus, 66123 Saarbr\"{u}cken, Germany}
\author{Luke C. G. Govia}
\altaffiliation[Current Address: ]{Department of Physics, McGill University, Montreal, Quebec, Canada H3A 2T8}
\affiliation{Theoretical Physics, Saarland University, Campus, 66123 Saarbr\"{u}cken, Germany}
\author{Frank K. Wilhelm}
\affiliation{Theoretical Physics, Saarland University, Campus, 66123 Saarbr\"{u}cken, Germany}

\begin{abstract}
Quantum walks are a promising framework that can be used to both understand and implement quantum information processing tasks. The quantum stochastic walk is a recently developed framework that combines the concept of a quantum walk with that of a classical random walk, through open system evolution of a quantum system. Quantum stochastic walks have been shown to have applications in as far reaching fields as artificial intelligence. However, there are significant constraints on the kind of open system evolutions that can be realized in a physical experiment. In this work, we discuss the restrictions on the allowed open system evolution, and the physical assumptions underpinning them. We show that general implementations would require the complete solution of the underlying unitary dynamics, and sophisticated reservoir engineering, thus weakening the benefits of experimental investigations.
\end{abstract}

\maketitle
With the ever increasing experimental control over single and complex quantum systems ~\cite{Jones2011,Devoret2013,Cho2015,Glaser2015}, harvesting the power of quantum physics for new technologies is no longer a far-fetched idea. For a clear example of the quantum world entering day-to-day life, one need look no further than quantum cryptography~\cite{Gisin:2002aa}. Quantum walks~\cite{Aharonov1993} are one of the most prominent frameworks in which to design and think about quantum algorithms. Both the continuous-~\cite{Farhi1998} and discrete-time~\cite{Aharonov2001,ambainis2001one} versions have been shown to provide speed-up over classical information processing tasks~\cite{Ambainis2007}, and are universal for quantum computing ~\cite{Childs2003,Lovett2010}. Classical (probabilistic) and quantum unitary random walks yield different distributions due to interference effects between different paths the walker can take on the associated graph network. Combining the two, a stochastic, continuous-time quantum walk (QSW) can be defined in an axiomatic manner to include unitary and non-unitary effects, and include both classical and quantum walks as limiting cases~\cite{Whitfield2010}.

While a general purpose quantum computer is still far over the horizon, intermediary technologies have been emerging with the promise to breach classical limitations. Within these, implementations of quantum neural networks, efficient quantum transport, and boson sampling have appeared as some of the platforms displaying the power of quantum walks. In many cases, such as excitation transfer in photosynthetic complexes, or in neural networks, one of the key questions is the role of coherence in the process efficiency. Therefore, their description in a QSW formalism is natural~\cite{Schuld2014c,Mohseni2008}. More recently, novel exciting applications have been proposed in artificial intelligence~\cite{Rebentrost2014,Cuevas,Briegel}, which, at its core, involves an autonomous agent that can learn from environmental input and react to it, changing its behavior as more input is received. A recent proposal uses quantum stochastic walks to speed-up the learning of the agent~\cite{Cuevas}.

As a standard quantum walk arises from unitary evolution, which is a special case of reversible evolution, the associated graphs are undirected. However, the stochastic processes present in QSWs can give some directionality to the graph network, at the price of introducing decoherence. Directionality, in turn, can enhance transport~\cite{Zimboras2013a}, or speed-up memory access in artificial intelligence~\cite{Cuevas}. In order to preserve quantum speedup, the nature of decoherence, i.e. its selectivity and rate, need to be carefully designed.

In this work we investigate the implementation of QSWs with no active control of the environment or ancillary systems. We describe the limitations to physical implementations of such a QSW, and show that only a very restricted set of graphs can be implemented with quantum systems under the canonical nondegenerate weak-coupling assumptions. Our results suggest that a large class of master equations often found in literature cannot be engineered in real systems without either active environmental control, or solving the full system dynamics prior to its implementation.

The classical random walk is an important and well studied model in statistical physics \cite{vanKampen97} describing the probabilistic movement of a walker along a graph. System evolution in a classical random walk is described by the transition probabilities from one node of the graph to all other nodes connected to it. In a quantum walk the evolution is instead described by transition probability amplitudes, that is, by unitary evolution of the quantum state describing the walker. Therefore, the main difference between classical and quantum walks is coherence.

The generalization encompassing both concepts is the so called quantum stochastic walk \cite{Whitfield2010}. Each quantum stochastic walk is defined by its Lindblad master equation, which can be generally written as
\begin{align}
\label{eq:QSW_Lindblad}
\dot{\rho}=-i\left[\hat{H},\rho\right]+\sum_{k}\gamma_k{\left(\hat{L}_k\rho\hat{L}_k^{\dagger}-\frac{1}{2}\big\{\hat{L}_k^{\dagger}\hat{L}_k,\rho\big\}\right)},
\end{align}
where $\rho$ denotes the density operator of the walker, $\hat{L}_k$ are the Lindblad operators and $\gamma_k$ the associated rates (transition probabilities) responsible for the incoherent part of the time evolution, and $\hat{H}$ is the Hamiltonian describing the coherent part of the time evolution. The structure of the graph is encoded by the associated matrices - nonzero elements of $\hat{H}$ encode coherent edges and those of the Lindblad operators encode incoherent edges. Equation (\ref{eq:QSW_Lindblad}) incorporates both the classical random walk and the quantum walk as special cases, and also allows for the study of more general walks that exhibit both coherent quantum and random classical behavior.

The quantum stochastic walk has been used to develop new quantum algorithms, such as the machine learning algorithms of Refs.~\cite{Cuevas,Briegel}, and to model transport phenomena in photosynthetic complexes~\cite{Zimboras2013a} and quantum neural networks~\cite{Schuld2014c}. In these, the incoherent evolution conserves the excitation number of the graph, so that evolution remains in the single excitation subspace when a single walker is present, and the walker cannot be lost. If we consider the situation where each node of the graph is represented by a qubit, then the Lindblad operator describing incoherent excitation exchange from node $n$ to node $m$ of the graph is given by $\hat{L}_k = \sigma_n^{-}\sigma_m^{+}$. The full Lindblad equation is therefore given by
\begin{align}
&\nonumber\dot{\rho}=-i\left[\hat{H},\rho\right] \\ &+\sum_{nm}\gamma_{nm}{\left(\sigma_m^{+}\sigma_n^{-}\rho\sigma_n^{+}\sigma_m^{-}-\frac{1}{2}\big\{\sigma_n^{+}\sigma_m^{-}\sigma_m^{+}\sigma_n^{-},\rho\big\}\right)}, 
\label{eq:QSWs}
\end{align}
and it is the physical implementation of this walk that will be the focus of this work. Note that for a general, potentially directed walk, we must allow $\gamma_{nm}\not=\gamma_{mn}$.

The physical implementation of the quantum stochastic walk of eq.(\ref{eq:QSWs}) is challenging, as incoherent excitation exchange between the nodes is required, while at the same time the excitation must be protected from decaying into the environment. In the following we will show explicitly that the incoherent evolution of eq.(\ref{eq:QSWs}) cannot be built without active control of the system and environment, or the solution of the complete unitary dynamics. While it is well known that master equations built from phenomenological models can lead to unphysical results~\cite{Scala2007,Beaudoin2011}, our work, on the other hand, focuses on whether a microscopic implementation can be created to mimic the desired dynamics.

It is important to note that in this work we consider the quantum stochastic walk as defined in Ref.~\cite{Whitfield2010}, and not the open quantum walk of Refs.~\cite{Attal:2012aa,Sinayskiy:2013aa}, which, despite the similar nomenclature, is an entirely different framework of open system evolution involving both internal and positional degrees of freedom of the walker. The QSW investigated here follows the paradigm of Ref.~\cite{Farhi1998}, where the continuous time evolution takes place exclusively on the position space of the walker, with no internal (coin) space needed.

{\em Building a QSW from Standard Decoherence Models.} We consider the standard microscopic derivation of a Lindblad form master equation in the weak coupling limit \cite{Breuer:2006uq} using the secular approximation. While in principle the Lindblad equation can also be derived in the singular coupling limit \cite{Breuer:2006uq}, this situation only plays a minor role in quantum computing platforms as it requires either strong damping or high bath temperature for the Markov approximation to be valid. Even though in the case of a QSW this would only apply to {\em some} degrees of freedom, it is more than challenging to engineer a quantum system where some degrees of freedom are coherent and others are strongly damped in a controlled way. 

We assume from here on a system with nondegenerate transition frequencies. We begin with a Hamiltonian of the form
\begin{align}
\hat{H} = \hat{H}_{\rm S} + \hat{H}_{\rm B} + \hat{H}_{\rm SB},
\end{align}
where $\hat{H}_{\rm S}$ is the self-Hamiltonian of the system, which in our case describes the graph, and $\hat{H}_{\rm B}$ is the self-Hamiltonian of the environmental bath. The term $\hat{H}_{\rm SB}$ describes the system-bath interaction, and without loss of generality has the form
\begin{align}
\hat{H}_{\rm SB} = \sum_{k,j}\widetilde\eta_k^{(j)}\hat{S}_k^{(j)} \otimes \hat{B}_k,
\label{eq:sys-bath-coupling}
\end{align}
where $\hat{S}_k^{(j)}$ and $\hat{B}_k$ are system and bath operators respectively, and each bath $k$ can interact with many local nodes $j$ with a coupling strength given by $\widetilde\eta_k^{(j)}$. Starting from these, our goal is to obtain a Lindblad master equation of the form of eq.(\ref{eq:QSWs}). In the following, Latin letters are used for the system local basis states $\{\ket{m}\}$, and Greek letters for the system eigenbasis, $\{\ket{\alpha}\}$. In the local basis, the master equation takes the form~\cite{Garanin2011a}

\begin{align}
\frac{d}{dt}\rho _{mn}=&-\frac{i}{\hbar }\sum_{m^\prime}H_{s,mm^{\prime
}}\rho _{m^\prime n}+\frac{i}{\hbar }\sum_{n^\prime}\rho _{mn^{\prime
}}H_{s,n^\prime n}\nonumber\\
&+\sum_{m^\prime n^\prime}R_{mn,m^\prime n^{\prime}}\rho _{m^\prime n^\prime},
\label{eq:DME-local}
\end{align}
where $H_{s,mm^\prime}\equiv \bra{m}\hat{H}_{s}\ket{m^\prime}$ describes coherent dynamics and $R_{mn,m^\prime n^\prime}$ the incoherent transition rates in the local basis. 

Within the secular approximation, fast oscillating terms are neglected and the transition rates between diagonal elements, i.e., $R_{mm,nn}\equiv\Gamma_{mn}$, can be written as
\begin{align}
\Gamma_{mn}=\sum_{\alpha \neq\alpha^\prime}T_{mn;\alpha\alpha^\prime}\Gamma_{\alpha \alpha^\prime}
-\sum_{\alpha \beta}\widetilde T_{mn;\alpha\beta}\widetilde\Gamma_{\alpha \beta}.
\label{eq:Gamma_local}
\end{align}
Here $\Gamma_{\alpha\alpha ^\prime}$ represents the transition rate between eigenstates, $\widetilde\Gamma_{\alpha\beta}$ their total dephasing rate, and we have defined $\widetilde T_{mn;\alpha\beta}\equiv \braket{m}{\alpha}\braket{\beta}{m}\braket{\alpha}{n}\braket{n}{\beta}$ and $T_{mn;\alpha\alpha^\prime}\equiv \left|\braket{m}{\alpha}\right|^2\left|\braket{\alpha^\prime}{n}\right|^2$. All other contributions are suppressed under the secular approximation, as we have assumed all transition frequencies are nondegenerate.

As discussed before, eq.(\ref{eq:QSWs}) does not allow for incoherent annihilation of local excitations and so we require that $\Gamma_{mn}=0$ whenever $\ket{m}$ and $\ket{n}$ contain a different number of excitations. However, as $\Gamma_{\alpha\alpha^\prime},\,\widetilde\Gamma_{\alpha\beta},\,T_{mn;\alpha\alpha^\prime}\geq0$ the first sum in eq.(\ref{eq:Gamma_local}) will be non-negative. Vanishing transition rates then mean that either both sums vanish, or the second one is negative and the sums cancel each other.

{\em Eigenstate transition rates.} To determine the conditions required to set $\Gamma_{mn}=0$ we start by looking at the second sum in eq.(\ref{eq:Gamma_local}). Non-zero terms in the sum would require pairs of eigenstates $\{\ket\alpha,\ket\beta\}$ with non-zero overlap with both $\ket m$ and $\ket n$, otherwise $\widetilde T_{mn;\alpha\beta} = 0$. Thus, avoiding transitions between local states with different excitation numbers, by cancelling the two sums of eq.(\ref{eq:Gamma_local}), would demand eigenstates of the system Hamiltonian spanning states with different numbers of walkers. However, applications of quantum walks usually rely on number-conserving graphs, at least at the Hamiltonian level. As such, $\widetilde T_{mn;\alpha\beta} = 0$ for all $\ket{m}$ and $\ket{n}$ with a different number of excitations, and the second sum in eq.(\ref{eq:Gamma_local}) vanishes. Therefore, the first sum in eq.(\ref{eq:Gamma_local}) must also vanish so that $\Gamma_{mn}=0$.

The transition rates between eigenstates are given by~\cite{Garanin2011a}
\begin{align}
\Gamma _{\alpha \alpha ^\prime}=\frac{2\pi }{\hbar Z_{b}}\sum_{\omega \omega
^\prime}&e^{-E_{\omega ^\prime}/(k_{B}T)}\delta \left( \epsilon
_{\alpha }-\epsilon _{\alpha ^\prime}+E_{\omega }-E_{\omega ^{\prime
}}\right)\nonumber\\
&\times\left| V_{\alpha \omega ,\alpha ^\prime\omega ^{\prime
}}\right| ^{2},
\label{eq:Gamma_eigen}
\end{align}
where $Z_b$ is the bath partition function, $k_B$ the Boltzmann constant, and $T$ the bath temperature. Here $\epsilon_\alpha$ and $\ket\alpha$ ($E_\omega$ and $\ket{\phi_B(\omega)}$) are the system (bath) eigenenergies and eigenvectors, and
\begin{align}
V_{\alpha\omega ,\alpha^\prime\omega^\prime}\equiv \bra{\alpha,\phi_B(\omega)}
\hat{H}_{SB}\ket{\alpha^\prime,\phi_B(\omega^\prime)}.
\end{align}
For the interaction Hamiltonian of eq.(\ref{eq:sys-bath-coupling}) we have
\begin{align}
V_{\alpha\omega ,\alpha^\prime\omega^\prime}&=\bra{\alpha,\phi_B(\omega)}\sum_{k,j}\widetilde\eta_k^{(j)}\hat{S}_k^{(j)} \otimes \hat{B}_k\ket{\alpha^\prime,\phi_B(\omega^\prime)}\nonumber\\
&=\sum_{k,j}\eta_k^{(j)}\bra{\alpha}\hat{S}_k^{(j)}\ket{\alpha^\prime}\nonumber\\
&\equiv\bra{\alpha}\overline V(\omega,\omega^\prime)\ket{\alpha^\prime}.
\label{eq:sei-la}
\end{align}

Let us focus on transitions between states $\ket m$ and $\ket n$ with
\begin{align}
\ket{m}=\sum_\alpha c_\alpha\ket\alpha\quad;\quad
\ket{n}=\sum_{\alpha^\prime} c_{\alpha^\prime}^\prime\ket{\alpha^\prime}.
\end{align}
The first sum in eq.(\ref{eq:Gamma_local}) vanishes only if $\Gamma_{\alpha\alpha^\prime}=0$ for all $\{\alpha,\alpha^\prime\}$ with $c_\alpha,\,c_{\alpha^\prime}^\prime\neq0$. From eq.(\ref{eq:Gamma_eigen}) we see that $\Gamma_{\alpha\alpha^\prime}=0$ only if $V_{\alpha\omega ,\alpha^\prime\omega^\prime}$ vanishes for all bath states for which the Dirac Delta is non-zero. Let us assume that the bath spectrum is dense, as is already implied by the Markov approximation being applicable, such that the energy conservation condition $\epsilon
_{\alpha }-\epsilon _{\alpha ^\prime}+E_{\omega }-E_{\omega ^{\prime}} = 0$ can always be fulfilled. This means that, to prevent transitions $\ket{m} \leftrightarrow\ket{n}$ we require that for all $\{\alpha,\alpha^\prime\}$ with $c_\alpha,\,c_{\alpha^\prime}^\prime\neq0$, the following statements can be inferred from one another
\begin{align}
&\bra\alpha\overline{V}(\omega,\omega^\prime)\ket{\alpha^\prime}=0\\
\implies &c_\alpha c_{\alpha^\prime}^\prime\bra\alpha\overline{V}(\omega,\omega^\prime)\ket{\alpha^\prime}=0\\
\implies &\sum_{\alpha,\alpha^\prime}c_\alpha c_{\alpha^\prime}^\prime\bra\alpha\overline{V}(\omega,\omega^\prime)\ket{\alpha^\prime}=0\\
\implies &\bra{m}\overline{V}(\omega,\omega^\prime)\ket{n}=0.
\label{eq:V_mn}
\end{align}
Of course, the validity of this last equality does not imply $\Gamma_{\alpha\alpha^\prime}=0$, but if it does not hold, then $\Gamma_{\alpha\alpha^\prime}\neq0$ for at least one pair $\{\alpha,\alpha^\prime\}$.

{\em Semi-local decay into the reservoir.} Let us now look at the case where $\hat{S}_k^{(j)}=\hat\sigma_x^{(j)}$ and work out the above condition, eq.(\ref{eq:V_mn}), for $\ket{n}=\ket 1_n$ (one walker at node $n$) and $\ket{m}=\ket 0$ (vacuum state). For this,
\begin{align}
\bra{m}\overline{V}(\omega,\omega^\prime)\ket{n}&=\sum_{k,j}\eta_k^{(j)}\bra{0}\hat\sigma_x^{(j)}\ket{1}_n=\sum_{k}\eta_k^{(n)}=0.
\end{align}
As all transitions $\ket 1_n\leftrightarrow\ket0$ have to be avoided, this must hold for all $n$, and so
\begin{align}
\overline{V}(\omega,\omega^\prime)&=\sum_{k,j}\eta_k^{(j)}\hat\sigma_x^{(j)}=\sum_j\hat\sigma_x^{(j)}\sum_k\eta_k^{(j)}=0.
\label{eq:V_sigma_x}
\end{align}
This is an effective bath-system decoupling, as $\bra{\alpha}\overline V(\omega,\omega^\prime)\ket{\alpha^\prime} = 0$, which implies $\Gamma_{\alpha\alpha'} = 0$ for all eigenstates $\ket{\alpha}, \ \ket{\alpha'}$. Therefore, if we require that the first sum in eq.\eqref{eq:Gamma_local} vanish for all transitions $\ket 1_n\leftrightarrow\ket0$, it must vanish for all transitions, regardless of the nodes and number of excitations involved.

As the second sum in eq.(\ref{eq:Gamma_local}) vanishes for standard QWs,  and as we have just shown, so too must the first, we argue that local-decay cannot be used to generate any relevant dynamics of the form of eq.(\ref{eq:QSWs}). We note that it is clear that the calculations shown above can be directly generalized to $\hat S_k^{(j)}=\hat\sigma_y^{(j)}$.

{\em Semi-local reservoir induced dephasing.} We now look at $\hat{S}_k^{(j)}=\hat\sigma_z^{(j)}$. From the arguments above, we immediately see that this can in principle be used to generate $\Gamma_{\alpha\alpha^\prime}=0$, as 
\begin{align}
\sum_{k,j}\eta_k^{(j)}\bra{m}\hat\sigma_z^{(j)}\ket{n}=0
\end{align}
is trivially fulfilled if $\ket{m}$ and $\ket{n}$ have different local excitation numbers. By the same token, in the 1-walker subspace, this is also fulfilled by any two states $\ket1_m$ and $\ket1_n$, with $m\neq n$. However, this does not imply $\Gamma_{mn}=0$, as mentioned before. If, for example, two eigenstates spam both $\ket1_m$ and $\ket1_n$, then it is possible to have $\Gamma_{mn}\neq0$.

The next point is to see if we can cancel local pure-dephasing, while maintaining $\Gamma_{mn}\neq0$ between single excitation states of the nodes. Local pure-dephasing is described by the rate
\begin{align}
\widetilde\Gamma_{mn}\equiv R_{mn,mn} =\sum_{\alpha \neq\alpha^\prime}\widetilde T_{mn;\alpha\alpha^\prime}\Gamma_{\alpha \alpha^\prime}
-\sum_{\alpha \beta} T_{mn;\alpha\beta}\widetilde\Gamma_{\alpha \beta},
\end{align}
and we desire that this vanish. As we require that $\Gamma_{mn}\neq0$, there must be eigenstates for which $ \widetilde T_{mn;\alpha\alpha^\prime},\ \Gamma_{\alpha \alpha^\prime}, \ T_{mn;\alpha\beta},\ \widetilde \Gamma_{\alpha \beta} \neq 0$. Therefore, in general, we need to cancel the two sums to obtain $\widetilde\Gamma_{mn}=0$, which requires an environment tailored to the graph. While engineering an environment that is locally compatible with the graph Hamiltonian is in principle doable, adapting the environment to the global eigenstates of the graph is equivalent to first solving the problem the quantum computer is supposed to solve.

What we have found is that if the Hamiltonian of the graph preservers excitation number, then local dephasing can be used to obtain $\Gamma_{mn} \neq 0$ only for transitions that conserve excitation number, as desired. However, to do so we must first calculate the eigensystem of $H_S$, then solve a set of coupled equations that give the appropriate system-bath couplings for zero local pure-dephasing. This is effectively reservoir engineering, and can be a considerable undertaking when the graph is of sufficient size. However, it must be noted that as both dephasing and $H_S$ preserve excitation number, only one block of the Hamiltonian must be diagonalized. This is not true for local decay, which does not conserve excitation number.

{\em Depolarizing channel.} A common decoherence model is the depolarizing channel, described by a coupling of the reservoir to all qubit operators, $\sigma_x^{(j)}$, $\sigma_y^{(j)}$ and $\sigma_z^{(j)}$. The calculations above can be easily generalized to show that such coupling to the bath leads to all the restrictions found for both decay and dephasing. As expected, this model cannot lead to the desired master equation.

In both the decay and dephasing cases, it is in principle possible to solve the full set of equations, eq.(\ref{eq:Gamma_local}), and find parameters for which all the unwanted transitions vanish. However, this would involve full diagonalisation of both $\hat H_S$ and $\hat H_B$, and possibly (definitely in the case of local decay) require very intricate bath engineering. In addition, such techniques would intricately link the coherent and incoherent transition rates.

One possibility would be to design reservoirs such that for certain pairs of states, the energy conservation condition of eq.(\ref{eq:Gamma_eigen}) cannot be fulfilled, and thus, eq.(\ref{eq:V_mn}) would not be valid. In doing so, one would cancel unwanted transitions by a careful designed reservoir spectrum. While such strategies for reservoir engineering can be useful~\cite{Weimer2010,Murch2012,Fogarty2013,Govia:2015aa,Mourokh2015}, for the applications usually envisioned for quantum walks their use would rely on the laborious solution of the complete unitary problem.

We note that other types of Lindblad equations can be conceived, where local excitation decay plays an important role, e.g., in an energy transfer process~\cite{Mohseni2008}. However, as these models do not usually include incoherent creation of excitations, implementations of such dynamics would be plagued by the same issues discussed here. As a consequence of the secular approximation the rates $\Gamma_{mn}$ and $\Gamma_{nm}$ cannot be independently set, so preventing one such transition while allowing the other would pose the same requirements as for eq.(\ref{eq:QSWs}).

{\em Conclusions.} Master equations in Lindblad form describe the most general quantum state evolution that is guaranteed to be completely positive and trace preserving. However, the set of Lindblad operators allowed for a given physical system is limited by the physical interactions naturally occurring. In other words, often the mathematical formulation of a dynamical system cannot be realizable in real world applications. We have studied such limitations for an arbitrary, number-conserving, stochastic master equation, under the usual secular and Markov approximations. We have discussed the problems of creating such an evolution using ubiquitous decoherence models, such as pure dephasing, amplitude damping, and depolarizing channels.

Our results show that microscopic implementation of general open system evolution can only be realized if the full unitary dynamics have been solved, and control of the reservoir is available. For interesting cases, actual experimental implementations of quantum stochastic walks are intended to tackle classically hard problems, which makes solving the full unitary dynamics of the system infeasible. Moreover, as all two-body system-bath interactions can be described by the interaction Hamiltonians studied above, carefully designed interactions can only circumvent the restrictions found by properly engineered local reservoir spectra, or for systems for which the secular approximation does not apply. In the latter case, this could be done by suitable use of degenerate transitions; however, engineering these would again require complete knowledge of the unitary dynamics.

\begin{acknowledgments}
We thank P.K. Schuhmacher and R. Betzholz for fruitful discussions. Supported by the Army Research Office under contract W911NF-14-1-0080 and the European Union through ScaleQIT. LCGG acknowledges support from NSERC through an NSERC PGS-D.
\end{acknowledgments}

\bibliography{Bib_QSW_Realizability}

\begin{thebibliography}{31}
\expandafter\ifx\csname natexlab\endcsname\relax\def\natexlab#1{#1}\fi
\expandafter\ifx\csname bibnamefont\endcsname\relax
  \def\bibnamefont#1{#1}\fi
\expandafter\ifx\csname bibfnamefont\endcsname\relax
  \def\bibfnamefont#1{#1}\fi
\expandafter\ifx\csname citenamefont\endcsname\relax
  \def\citenamefont#1{#1}\fi
\expandafter\ifx\csname url\endcsname\relax
  \def\url#1{\texttt{#1}}\fi
\expandafter\ifx\csname urlprefix\endcsname\relax\def\urlprefix{URL }\fi
\providecommand{\bibinfo}[2]{#2}
\providecommand{\eprint}[2][]{\url{#2}}

\bibitem[{\citenamefont{Jones}(2011)}]{Jones2011}
\bibinfo{author}{\bibfnamefont{J.~A.} \bibnamefont{Jones}},
  \bibinfo{journal}{Progress in Nuclear Magnetic Resonance Spectroscopy}
  \textbf{\bibinfo{volume}{59}}, \bibinfo{pages}{91 } (\bibinfo{year}{2011}),
  ISSN \bibinfo{issn}{0079-6565},
  \urlprefix\url{http://www.sciencedirect.com/science/article/pii/S0079656510001111}.

\bibitem[{\citenamefont{Devoret and Schoelkopf}(2013)}]{Devoret2013}
\bibinfo{author}{\bibfnamefont{M.~H.} \bibnamefont{Devoret}} \bibnamefont{and}
  \bibinfo{author}{\bibfnamefont{R.~J.} \bibnamefont{Schoelkopf}},
  \bibinfo{journal}{Science} \textbf{\bibinfo{volume}{339}},
  \bibinfo{pages}{1169} (\bibinfo{year}{2013}), ISSN \bibinfo{issn}{0036-8075},
  \urlprefix\url{http://science.sciencemag.org/content/339/6124/1169}.

\bibitem[{\citenamefont{Cho et~al.}(2015)\citenamefont{Cho, Hong, Lee, and
  Kim}}]{Cho2015}
\bibinfo{author}{\bibfnamefont{D.-I.~D.} \bibnamefont{Cho}},
  \bibinfo{author}{\bibfnamefont{S.}~\bibnamefont{Hong}},
  \bibinfo{author}{\bibfnamefont{M.}~\bibnamefont{Lee}}, \bibnamefont{and}
  \bibinfo{author}{\bibfnamefont{T.}~\bibnamefont{Kim}},
  \bibinfo{journal}{Micro and Nano Systems Letters}
  \textbf{\bibinfo{volume}{3}}, \bibinfo{pages}{1} (\bibinfo{year}{2015}), ISSN
  \bibinfo{issn}{2213-9621},
  \urlprefix\url{http://dx.doi.org/10.1186/s40486-015-0013-3}.

\bibitem[{\citenamefont{Glaser et~al.}(2015)\citenamefont{Glaser, Boscain,
  Calarco, Koch, K{\"{o}}ckenberger, Kosloff, Kuprov, Luy, Schirmer,
  Schulte-Herbr{\"{u}}ggen et~al.}}]{Glaser2015}
\bibinfo{author}{\bibfnamefont{S.~J.} \bibnamefont{Glaser}},
  \bibinfo{author}{\bibfnamefont{U.}~\bibnamefont{Boscain}},
  \bibinfo{author}{\bibfnamefont{T.}~\bibnamefont{Calarco}},
  \bibinfo{author}{\bibfnamefont{C.~P.} \bibnamefont{Koch}},
  \bibinfo{author}{\bibfnamefont{W.}~\bibnamefont{K{\"{o}}ckenberger}},
  \bibinfo{author}{\bibfnamefont{R.}~\bibnamefont{Kosloff}},
  \bibinfo{author}{\bibfnamefont{I.}~\bibnamefont{Kuprov}},
  \bibinfo{author}{\bibfnamefont{B.}~\bibnamefont{Luy}},
  \bibinfo{author}{\bibfnamefont{S.}~\bibnamefont{Schirmer}},
  \bibinfo{author}{\bibfnamefont{T.}~\bibnamefont{Schulte-Herbr{\"{u}}ggen}},
  \bibnamefont{et~al.}, \bibinfo{journal}{The European Physical Journal D}
  \textbf{\bibinfo{volume}{69}}, \bibinfo{pages}{279} (\bibinfo{year}{2015}),
  \eprint{1508.00442},
  \urlprefix\url{http://link.springer.com/article/10.1140%2Fepjd%2Fe2015-60464-1}.

\bibitem[{\citenamefont{Gisin et~al.}(2002)\citenamefont{Gisin, Ribordy,
  Tittel, and Zbinden}}]{Gisin:2002aa}
\bibinfo{author}{\bibfnamefont{N.}~\bibnamefont{Gisin}},
  \bibinfo{author}{\bibfnamefont{G.}~\bibnamefont{Ribordy}},
  \bibinfo{author}{\bibfnamefont{W.}~\bibnamefont{Tittel}}, \bibnamefont{and}
  \bibinfo{author}{\bibfnamefont{H.}~\bibnamefont{Zbinden}},
  \bibinfo{journal}{Review of Modern Physics} \textbf{\bibinfo{volume}{74}},
  \bibinfo{pages}{145} (\bibinfo{year}{2002}),
  \urlprefix\url{http://journals.aps.org/rmp/abstract/10.1103/RevModPhys.74.145}.

\bibitem[{\citenamefont{Aharonov et~al.}(1993)\citenamefont{Aharonov,
  Davidovich, and Zagury}}]{Aharonov1993}
\bibinfo{author}{\bibfnamefont{Y.}~\bibnamefont{Aharonov}},
  \bibinfo{author}{\bibfnamefont{L.}~\bibnamefont{Davidovich}},
  \bibnamefont{and} \bibinfo{author}{\bibfnamefont{N.}~\bibnamefont{Zagury}},
  \bibinfo{journal}{Physical Review A} \textbf{\bibinfo{volume}{48}},
  \bibinfo{pages}{1687} (\bibinfo{year}{1993}),
  \urlprefix\url{http://link.aps.org/doi/10.1103/PhysRevA.48.1687}.

\bibitem[{\citenamefont{Farhi and Gutmann}(1998)}]{Farhi1998}
\bibinfo{author}{\bibfnamefont{E.}~\bibnamefont{Farhi}} \bibnamefont{and}
  \bibinfo{author}{\bibfnamefont{S.}~\bibnamefont{Gutmann}},
  \bibinfo{journal}{Physical Review A} \textbf{\bibinfo{volume}{58}},
  \bibinfo{pages}{915} (\bibinfo{year}{1998}), ISSN \bibinfo{issn}{1050-2947},
  \urlprefix\url{http://link.aps.org/doi/10.1103/PhysRevA.58.915}.

\bibitem[{\citenamefont{Aharonov et~al.}(2001)\citenamefont{Aharonov, Ambainis,
  Kempe, and Vazirani}}]{Aharonov2001}
\bibinfo{author}{\bibfnamefont{D.}~\bibnamefont{Aharonov}},
  \bibinfo{author}{\bibfnamefont{A.}~\bibnamefont{Ambainis}},
  \bibinfo{author}{\bibfnamefont{J.}~\bibnamefont{Kempe}}, \bibnamefont{and}
  \bibinfo{author}{\bibfnamefont{U.}~\bibnamefont{Vazirani}}, in
  \emph{\bibinfo{booktitle}{Proceedings of the thirty-third annual ACM
  symposium on Theory of computing - STOC '01}} (\bibinfo{publisher}{ACM
  Press}, \bibinfo{address}{New York, New York, USA}, \bibinfo{year}{2001}),
  pp. \bibinfo{pages}{50--59}, ISBN \bibinfo{isbn}{1581133499},
  \urlprefix\url{http://dl.acm.org/citation.cfm?id=380752.380758}.

\bibitem[{\citenamefont{Ambainis et~al.}(2001)\citenamefont{Ambainis, Bach,
  Nayak, Vishwanath, and Watrous}}]{ambainis2001one}
\bibinfo{author}{\bibfnamefont{A.}~\bibnamefont{Ambainis}},
  \bibinfo{author}{\bibfnamefont{E.}~\bibnamefont{Bach}},
  \bibinfo{author}{\bibfnamefont{A.}~\bibnamefont{Nayak}},
  \bibinfo{author}{\bibfnamefont{A.}~\bibnamefont{Vishwanath}},
  \bibnamefont{and} \bibinfo{author}{\bibfnamefont{J.}~\bibnamefont{Watrous}},
  in \emph{\bibinfo{booktitle}{Proceedings of the thirty-third annual ACM
  symposium on Theory of computing}} (\bibinfo{organization}{ACM},
  \bibinfo{year}{2001}), pp. \bibinfo{pages}{37--49},
  \urlprefix\url{http://dl.acm.org/citation.cfm?id=380757}.

\bibitem[{\citenamefont{Ambainis}(2007)}]{Ambainis2007}
\bibinfo{author}{\bibfnamefont{A.}~\bibnamefont{Ambainis}},
  \bibinfo{journal}{SIAM Journal on Computing} \textbf{\bibinfo{volume}{37}},
  \bibinfo{pages}{210} (\bibinfo{year}{2007}), ISSN \bibinfo{issn}{0097-5397},
  \urlprefix\url{http://epubs.siam.org/doi/abs/10.1137/S0097539705447311}.

\bibitem[{\citenamefont{Childs et~al.}(2003)\citenamefont{Childs, Cleve,
  Deotto, Farhi, Gutmann, and Spielman}}]{Childs2003}
\bibinfo{author}{\bibfnamefont{A.~M.} \bibnamefont{Childs}},
  \bibinfo{author}{\bibfnamefont{R.}~\bibnamefont{Cleve}},
  \bibinfo{author}{\bibfnamefont{E.}~\bibnamefont{Deotto}},
  \bibinfo{author}{\bibfnamefont{E.}~\bibnamefont{Farhi}},
  \bibinfo{author}{\bibfnamefont{S.}~\bibnamefont{Gutmann}}, \bibnamefont{and}
  \bibinfo{author}{\bibfnamefont{D.~A.} \bibnamefont{Spielman}}, in
  \emph{\bibinfo{booktitle}{Proceedings of the thirty-fifth ACM symposium on
  Theory of computing - STOC '03}} (\bibinfo{publisher}{ACM Press},
  \bibinfo{address}{New York, New York, USA}, \bibinfo{year}{2003}),
  p.~\bibinfo{pages}{59}, ISBN \bibinfo{isbn}{1581136749},
  \urlprefix\url{http://dl.acm.org/citation.cfm?id=780542.780552}.

\bibitem[{\citenamefont{Lovett et~al.}(2010)\citenamefont{Lovett, Cooper,
  Everitt, Trevers, and Kendon}}]{Lovett2010}
\bibinfo{author}{\bibfnamefont{N.~B.} \bibnamefont{Lovett}},
  \bibinfo{author}{\bibfnamefont{S.}~\bibnamefont{Cooper}},
  \bibinfo{author}{\bibfnamefont{M.}~\bibnamefont{Everitt}},
  \bibinfo{author}{\bibfnamefont{M.}~\bibnamefont{Trevers}}, \bibnamefont{and}
  \bibinfo{author}{\bibfnamefont{V.}~\bibnamefont{Kendon}},
  \bibinfo{journal}{Physical Review A} \textbf{\bibinfo{volume}{81}},
  \bibinfo{pages}{042330} (\bibinfo{year}{2010}),
  \urlprefix\url{http://journals.aps.org/pra/abstract/10.1103/PhysRevA.81.042330}.

\bibitem[{\citenamefont{Whitfield et~al.}(2010)\citenamefont{Whitfield,
  Rodr{\'{\i}}guez-Rosario, and Aspuru-Guzik}}]{Whitfield2010}
\bibinfo{author}{\bibfnamefont{J.~D.} \bibnamefont{Whitfield}},
  \bibinfo{author}{\bibfnamefont{C.~A.}
  \bibnamefont{Rodr{\'{\i}}guez-Rosario}}, \bibnamefont{and}
  \bibinfo{author}{\bibfnamefont{A.}~\bibnamefont{Aspuru-Guzik}},
  \bibinfo{journal}{Physical Review A} \textbf{\bibinfo{volume}{81}},
  \bibinfo{pages}{022323} (\bibinfo{year}{2010}), ISSN
  \bibinfo{issn}{1050-2947},
  \urlprefix\url{http://link.aps.org/doi/10.1103/PhysRevA.81.022323}.

\bibitem[{\citenamefont{Schuld et~al.}(2014)\citenamefont{Schuld, Sinayskiy,
  and Petruccione}}]{Schuld2014c}
\bibinfo{author}{\bibfnamefont{M.}~\bibnamefont{Schuld}},
  \bibinfo{author}{\bibfnamefont{I.}~\bibnamefont{Sinayskiy}},
  \bibnamefont{and}
  \bibinfo{author}{\bibfnamefont{F.}~\bibnamefont{Petruccione}},
  \bibinfo{journal}{Physical Review A} \textbf{\bibinfo{volume}{89}},
  \bibinfo{pages}{032333} (\bibinfo{year}{2014}), ISSN
  \bibinfo{issn}{1050-2947}, \eprint{1404.0159},
  \urlprefix\url{http://journals.aps.org/pra/abstract/10.1103/PhysRevA.89.032333}.

\bibitem[{\citenamefont{Mohseni et~al.}(2008)\citenamefont{Mohseni, Rebentrost,
  Lloyd, and Aspuru-Guzik}}]{Mohseni2008}
\bibinfo{author}{\bibfnamefont{M.}~\bibnamefont{Mohseni}},
  \bibinfo{author}{\bibfnamefont{P.}~\bibnamefont{Rebentrost}},
  \bibinfo{author}{\bibfnamefont{S.}~\bibnamefont{Lloyd}}, \bibnamefont{and}
  \bibinfo{author}{\bibfnamefont{A.}~\bibnamefont{Aspuru-Guzik}},
  \bibinfo{journal}{The Journal of Chemical Physics}
  \textbf{\bibinfo{volume}{129}}, \bibinfo{pages}{174106}
  (\bibinfo{year}{2008}), ISSN \bibinfo{issn}{1089-7690}, \eprint{0805.2741},
  \urlprefix\url{http://scitation.aip.org/content/aip/journal/jcp/129/17/10.1063/1.3002335}.

\bibitem[{\citenamefont{Rebentrost et~al.}(2014)\citenamefont{Rebentrost,
  Mohseni, and Lloyd}}]{Rebentrost2014}
\bibinfo{author}{\bibfnamefont{P.}~\bibnamefont{Rebentrost}},
  \bibinfo{author}{\bibfnamefont{M.}~\bibnamefont{Mohseni}}, \bibnamefont{and}
  \bibinfo{author}{\bibfnamefont{S.}~\bibnamefont{Lloyd}},
  \bibinfo{journal}{Physical Review Letters} \textbf{\bibinfo{volume}{113}},
  \bibinfo{pages}{130503} (\bibinfo{year}{2014}),
  \urlprefix\url{http://link.aps.org/doi/10.1103/PhysRevLett.113.130503}.

\bibitem[{\citenamefont{Briegel and las Cuevas}(2012)}]{Cuevas}
\bibinfo{author}{\bibfnamefont{H.~J.} \bibnamefont{Briegel}} \bibnamefont{and}
  \bibinfo{author}{\bibfnamefont{G.}~\bibnamefont{las Cuevas}},
  \bibinfo{journal}{Scientific Reports} \textbf{\bibinfo{volume}{2}},
  \bibinfo{pages}{400} (\bibinfo{year}{2012}),
  \urlprefix\url{http://www.nature.com/articles/srep00400}.

\bibitem[{\citenamefont{Paparo et~al.}(2014)\citenamefont{Paparo, Dunjko,
  Makmal, Martin-Delgado, and Briegel}}]{Briegel}
\bibinfo{author}{\bibfnamefont{G.~D.} \bibnamefont{Paparo}},
  \bibinfo{author}{\bibfnamefont{V.}~\bibnamefont{Dunjko}},
  \bibinfo{author}{\bibfnamefont{A.}~\bibnamefont{Makmal}},
  \bibinfo{author}{\bibfnamefont{M.~A.} \bibnamefont{Martin-Delgado}},
  \bibnamefont{and} \bibinfo{author}{\bibfnamefont{H.~J.}
  \bibnamefont{Briegel}}, \bibinfo{journal}{Physical Review X}
  \textbf{\bibinfo{volume}{4}}, \bibinfo{pages}{31002} (\bibinfo{year}{2014}),
  \urlprefix\url{http://link.aps.org/doi/10.1103/PhysRevX.4.031002}.

\bibitem[{\citenamefont{Zimbor{\'{a}}s
  et~al.}(2013)\citenamefont{Zimbor{\'{a}}s, Faccin, K{\'{a}}d{\'{a}}r,
  Whitfield, Lanyon, and Biamonte}}]{Zimboras2013a}
\bibinfo{author}{\bibfnamefont{Z.}~\bibnamefont{Zimbor{\'{a}}s}},
  \bibinfo{author}{\bibfnamefont{M.}~\bibnamefont{Faccin}},
  \bibinfo{author}{\bibfnamefont{Z.}~\bibnamefont{K{\'{a}}d{\'{a}}r}},
  \bibinfo{author}{\bibfnamefont{J.~D.} \bibnamefont{Whitfield}},
  \bibinfo{author}{\bibfnamefont{B.~P.} \bibnamefont{Lanyon}},
  \bibnamefont{and} \bibinfo{author}{\bibfnamefont{J.}~\bibnamefont{Biamonte}},
  \bibinfo{journal}{Scientific reports} \textbf{\bibinfo{volume}{3}},
  \bibinfo{pages}{2361} (\bibinfo{year}{2013}), ISSN \bibinfo{issn}{2045-2322},
  \urlprefix\url{http://www.nature.com/srep/2013/130806/srep02361/full/srep02361.html}.

\bibitem[{\citenamefont{van Kampen}(1997)}]{vanKampen97}
\bibinfo{author}{\bibfnamefont{N.}~\bibnamefont{van Kampen}},
  \emph{\bibinfo{title}{Stochastic processes in physics and chemistry}}
  (\bibinfo{publisher}{Elsevier}, \bibinfo{address}{Amsterdam},
  \bibinfo{year}{1997}).

\bibitem[{\citenamefont{Scala et~al.}(2007)\citenamefont{Scala, Militello,
  Messina, Piilo, and Maniscalco}}]{Scala2007}
\bibinfo{author}{\bibfnamefont{M.}~\bibnamefont{Scala}},
  \bibinfo{author}{\bibfnamefont{B.}~\bibnamefont{Militello}},
  \bibinfo{author}{\bibfnamefont{A.}~\bibnamefont{Messina}},
  \bibinfo{author}{\bibfnamefont{J.}~\bibnamefont{Piilo}}, \bibnamefont{and}
  \bibinfo{author}{\bibfnamefont{S.}~\bibnamefont{Maniscalco}},
  \bibinfo{journal}{Physical Review A} \textbf{\bibinfo{volume}{75}},
  \bibinfo{pages}{013811} (\bibinfo{year}{2007}),
  \urlprefix\url{http://link.aps.org/doi/10.1103/PhysRevA.75.013811}.

\bibitem[{\citenamefont{Beaudoin et~al.}(2011)\citenamefont{Beaudoin, Gambetta,
  and Blais}}]{Beaudoin2011}
\bibinfo{author}{\bibfnamefont{F.}~\bibnamefont{Beaudoin}},
  \bibinfo{author}{\bibfnamefont{J.~M.} \bibnamefont{Gambetta}},
  \bibnamefont{and} \bibinfo{author}{\bibfnamefont{A.}~\bibnamefont{Blais}},
  \bibinfo{journal}{Physical Review A} \textbf{\bibinfo{volume}{84}},
  \bibinfo{pages}{043832} (\bibinfo{year}{2011}),
  \urlprefix\url{http://link.aps.org/doi/10.1103/PhysRevA.84.043832}.

\bibitem[{\citenamefont{Attal et~al.}(2012)\citenamefont{Attal, Petruccione,
  Sabot, and Sinayskiy}}]{Attal:2012aa}
\bibinfo{author}{\bibfnamefont{S.}~\bibnamefont{Attal}},
  \bibinfo{author}{\bibfnamefont{F.}~\bibnamefont{Petruccione}},
  \bibinfo{author}{\bibfnamefont{C.}~\bibnamefont{Sabot}}, \bibnamefont{and}
  \bibinfo{author}{\bibfnamefont{I.}~\bibnamefont{Sinayskiy}},
  \bibinfo{journal}{Journal of Statistical Physics}
  \textbf{\bibinfo{volume}{147}}, \bibinfo{pages}{832} (\bibinfo{year}{2012}),
  ISSN \bibinfo{issn}{0022-4715},
  \urlprefix\url{http://link.springer.com/article/10.1007%2Fs10955-012-0491-0}.

\bibitem[{\citenamefont{Sinayskiy and Petruccione}(2013)}]{Sinayskiy:2013aa}
\bibinfo{author}{\bibfnamefont{I.}~\bibnamefont{Sinayskiy}} \bibnamefont{and}
  \bibinfo{author}{\bibfnamefont{F.}~\bibnamefont{Petruccione}},
  \bibinfo{journal}{Journal of Physics: Conference Series}
  \textbf{\bibinfo{volume}{442}}, \bibinfo{pages}{012003}
  (\bibinfo{year}{2013}),
  \urlprefix\url{http://iopscience.iop.org/article/10.1088/1742-6596/442/1/012003/meta}.

\bibitem[{\citenamefont{Breuer and Petruccione}(2006)}]{Breuer:2006uq}
\bibinfo{author}{\bibfnamefont{H.-P.} \bibnamefont{Breuer}} \bibnamefont{and}
  \bibinfo{author}{\bibfnamefont{F.}~\bibnamefont{Petruccione}},
  \emph{\bibinfo{title}{The Theory of Open Quantum Systems}}
  (\bibinfo{publisher}{Oxford University Press}, \bibinfo{year}{2006}).

\bibitem[{\citenamefont{Garanin}(2011)}]{Garanin2011a}
\bibinfo{author}{\bibfnamefont{D.~A.} \bibnamefont{Garanin}},
  \emph{\bibinfo{title}{{Advances in Chemical Physics}}}, Advances in Chemical
  Physics (\bibinfo{publisher}{John Wiley {\&} Sons, Inc.},
  \bibinfo{address}{Hoboken, NJ, USA}, \bibinfo{year}{2011}), ISBN
  \bibinfo{isbn}{9781118135242}, \eprint{0805.0391},
  \urlprefix\url{http://onlinelibrary.wiley.com/doi/10.1002/9781118135242.ch4/summary}.

\bibitem[{\citenamefont{Weimer et~al.}(2010)\citenamefont{Weimer, Muller,
  Lesanovsky, Zoller, and Buchler}}]{Weimer2010}
\bibinfo{author}{\bibfnamefont{H.}~\bibnamefont{Weimer}},
  \bibinfo{author}{\bibfnamefont{M.}~\bibnamefont{Muller}},
  \bibinfo{author}{\bibfnamefont{I.}~\bibnamefont{Lesanovsky}},
  \bibinfo{author}{\bibfnamefont{P.}~\bibnamefont{Zoller}}, \bibnamefont{and}
  \bibinfo{author}{\bibfnamefont{H.~P.} \bibnamefont{Buchler}},
  \bibinfo{journal}{Nature Physics} \textbf{\bibinfo{volume}{6}},
  \bibinfo{pages}{382} (\bibinfo{year}{2010}), ISSN \bibinfo{issn}{1745-2473},
  \urlprefix\url{http://dx.doi.org/10.1038/nphys1614}.

\bibitem[{\citenamefont{Murch et~al.}(2012)\citenamefont{Murch, Vool, Zhou,
  Weber, Girvin, and Siddiqi}}]{Murch2012}
\bibinfo{author}{\bibfnamefont{K.~W.} \bibnamefont{Murch}},
  \bibinfo{author}{\bibfnamefont{U.}~\bibnamefont{Vool}},
  \bibinfo{author}{\bibfnamefont{D.}~\bibnamefont{Zhou}},
  \bibinfo{author}{\bibfnamefont{S.~J.} \bibnamefont{Weber}},
  \bibinfo{author}{\bibfnamefont{S.~M.} \bibnamefont{Girvin}},
  \bibnamefont{and} \bibinfo{author}{\bibfnamefont{I.}~\bibnamefont{Siddiqi}},
  \bibinfo{journal}{Physical Review Letters} \textbf{\bibinfo{volume}{109}},
  \bibinfo{pages}{183602} (\bibinfo{year}{2012}),
  \urlprefix\url{http://link.aps.org/doi/10.1103/PhysRevLett.109.183602}.

\bibitem[{\citenamefont{Fogarty et~al.}(2013)\citenamefont{Fogarty, Kajari,
  Taketani, Busch, and Morigi}}]{Fogarty2013}
\bibinfo{author}{\bibfnamefont{T.}~\bibnamefont{Fogarty}},
  \bibinfo{author}{\bibfnamefont{E.}~\bibnamefont{Kajari}},
  \bibinfo{author}{\bibfnamefont{B.~G.} \bibnamefont{Taketani}},
  \bibinfo{author}{\bibfnamefont{T.}~\bibnamefont{Busch}}, \bibnamefont{and}
  \bibinfo{author}{\bibfnamefont{G.}~\bibnamefont{Morigi}},
  \bibinfo{journal}{Physical Review A} \textbf{\bibinfo{volume}{87}},
  \bibinfo{pages}{050304} (\bibinfo{year}{2013}),
  \urlprefix\url{http://link.aps.org/doi/10.1103/PhysRevA.87.050304}.

\bibitem[{\citenamefont{Govia and Wilhelm}(2015)}]{Govia:2015aa}
\bibinfo{author}{\bibfnamefont{L.~C.~G.} \bibnamefont{Govia}} \bibnamefont{and}
  \bibinfo{author}{\bibfnamefont{F.~K.} \bibnamefont{Wilhelm}},
  \bibinfo{journal}{Physical Review Applied} \textbf{\bibinfo{volume}{4}},
  \bibinfo{pages}{054001} (\bibinfo{year}{2015}),
  \urlprefix\url{http://link.aps.org/doi/10.1103/PhysRevApplied.4.054001}.

\bibitem[{\citenamefont{Mourokh and Nori}(2015)}]{Mourokh2015}
\bibinfo{author}{\bibfnamefont{L.~G.} \bibnamefont{Mourokh}} \bibnamefont{and}
  \bibinfo{author}{\bibfnamefont{F.}~\bibnamefont{Nori}},
  \bibinfo{journal}{Phys. Rev. E} \textbf{\bibinfo{volume}{92}},
  \bibinfo{pages}{052720} (\bibinfo{year}{2015}),
  \urlprefix\url{http://link.aps.org/doi/10.1103/PhysRevE.92.052720}.

\end{thebibliography}

\end{document}